\documentclass[%
 prx,reprint,floatfix,a4paper,
 amsmath,amssymb,
 aps,
 longbibliography
]{revtex4-1}

\usepackage[braket, qm]{qcircuit}
\usepackage{graphicx}
\usepackage{dcolumn}
\usepackage{bm}
\usepackage{fixltx2e}
\usepackage[colorlinks=true,citecolor=blue,linkcolor=blue,urlcolor=magenta]{hyperref}
\usepackage{url}
\DeclareMathOperator{\Tr}{Tr}

\usepackage{color}

\DeclareFontFamily{OMX}{MnSymbolE}{}
\DeclareFontShape{OMX}{MnSymbolE}{m}{n}{
    <-6>  MnSymbolE5
   <6-7>  MnSymbolE6
   <7-8>  MnSymbolE7
   <8-9>  MnSymbolE8
   <9-10> MnSymbolE9
  <10-12> MnSymbolE10
  <12->   MnSymbolE12}{}
\DeclareSymbolFont{mnlargesymbols}{OMX}{MnSymbolE}{m}{n}
\SetSymbolFont{mnlargesymbols}{bold}{OMX}{MnSymbolE}{b}{n}
\DeclareMathDelimiter{\llangle}{\mathopen}{mnlargesymbols}{'164}{mnlargesymbols}{'164}
\DeclareMathDelimiter{\rrangle}{\mathclose}{mnlargesymbols}{'171}{mnlargesymbols}{'171}

\begin{document}
	

\title{Quantum leakage detection using a model-independent dimension witness}

\author{Armands Strikis}
\email{Armands.Strikis@gmail.com}
\author{Animesh Datta}
\email{animesh.datta@warwick.ac.uk}
\author{George~C.~Knee}
\email{gk@physics.org}
\address{Department of Physics, University of Warwick, Coventry CV4 7AL, United Kingdom}
\date{\today}

\begin{abstract}
Users of quantum computers must be able to confirm they are indeed functioning as intended, even when the devices are remotely accessed. In particular, if the Hilbert space dimension of the components are not as advertised -- for instance if the qubits suffer leakage -- errors can ensue and protocols may be rendered insecure.  We refine the method of delayed vectors, adapted from classical chaos theory to quantum systems, and apply it remotely on the IBMQ platform -- a quantum computer composed of transmon qubits.  The method witnesses, in a model-independent fashion, dynamical signatures of higher-dimensional processes. We present evidence, under mild assumptions, that the IBMQ transmons suffer state leakage, with a $p$ value no larger than $5{\times}10^{-4}$ under a single qubit operation. We also estimate the number of shots necessary for revealing leakage in a two-qubit system.
\end{abstract}

\maketitle


\section{Introduction}
\label{intro}
Computing is entering a new era of remotely-accessible quantum machines, with examples in optical~\cite{BristolCloud} and superconducting~\cite{IBM,Rigetti} platforms: stepping stones on the route to full scale, error-corrected quantum computers~\cite{AcinBlochBuhrman2018} promising dramatic speedups over their classical counterparts~\cite{FruchtmanChoi2016}. It is desirable to seek assurance on the reliable operation of such machines, even when the user has no direct physical access to the interior of the `server' to which a quantum computation has been delegated~\cite{KashefiPappa2017}. Furthermore, this must be possible even when the remote user is unable or unwilling to rely on particular details of the hardware platform. 

Characterisation and mitigation of the errors affecting quantum computer components is vital for their reliable operation. Recent years have seen great improvements in the performance of one- and two-qubit gate in superconducting~\cite{Barends2014,Sheldon2016} and trapped ion platforms~\cite{Ballance2016,Schfer2018}. Further improvements in gate performance is often limited by leakage -- the phenomenon of quantum information evolving outside of a predefined computational subspace (such as a qubit). Leakage is of particular concern for quantum computer engineers, due to its impact on fault-tolerance thresholds~\cite{SucharaCrossGambetta2015}. For remote users, it may undermine the assumption of trusted operations -- preparation, memory or measurement -- that quantum verification~\cite{Gheorghiu2018} or quantum key distribution protocols~\cite{Hanggi2010} rely on to guarantee their security. Leakage has been studied in some detail in specific quantum computing platforms~\cite{Merket2011,Medford2013,Mehl2015,Chen2016,WillschNoconJin2017}. General frameworks have been developed, but rely on certain assumptions. For example, that any population in the leakage space is depolarised and that twirling over Clifford gates averages out coherences between the computational and leakage subspaces~\cite{WoodGambetta2018}, as well as those assumptions concomitant with randomised benchmarking~\cite{WallmanBarnhillEmerson2016}. For a distant user seeking to ascertain the reliability of a remote quantum device, the invalidity of these assumptions, as well as the specific laboratory operations which the frameworks call for,  is a severe difficulty. We overcome these issues by taking a different approach: Leakage may be detected by witnessing the existence of a process operating in higher dimensions than expected.

 Dimension witnesses (DWs) are functions of observable probabilities exhibiting dimension-dependent bounds, and were originally based on entanglement witnesses such as the Bell or CHSH inequality~\cite{BrunnerPironioAcin2008}. A DW has also been developed from a quantum coherence witness~\cite{SchildEmary2015}. All these approaches require specific operations to be faithfully implemented --- a major drawback. Moreover, many states will fail to trigger such DWs despite having ostensibly sufficient dimension. DWs based on nonlocal correlations are inapplicable to systems lacking a tensor-product structure. For the coherence witness-based DW~\cite{SchildEmary2015}, the dimension dependence becomes weaker with increasing dimension: implying its diminishing robustness to noise. 

So-called device-independent DWs have also been developed~\cite{WehnerChristandlDoherty2008,GallegoBrunnerHadley2010,HoffmannSpeeGuhne2018} and tested~\cite{AhrensBadziagCabello2012,AguilarFarkasMartinez2018}. Device-independent DWs do not rely on the faithful implementation of any particular preparations or measurements for their validity, but instead on reasoning about the space of conditional probabilities describing a set of prepare-and-measure experiments. Drawbacks to these approaches include (i) that the DWs themselves can be difficult to find, e.g. sometimes being the output of heuristic numerical routines; and (ii) states and measurements must be carefully engineered in order to trigger the witnesses, and experiments therefore need to be very precise in order to be successful. These are impediments to remote users as well as designers of quantum computers.

The method of delays originated as a tool to extract the dimension of classical chaotic flows from time series~\cite{Packard1980,Froehling1981} and was adapted to quantum mechanical systems by Wolf and Perez-Garcia (WP-G) in 2009~\cite{WolfPerez-Garcia2009}. It bounds the dimension of a system by analysing only the time series emanating from it, and is a more transparent and flexible approach to platform- and model-independent dimension witnessing. It makes a minimum of assumptions, both with respect to the preparations and measurements available, as well as with respect to the dynamics of the system. Moreover, it is relatively easy to find an experimental prescription that will expose the dimensionality of the system at hand.

In terms of quantum computation, witnessing a process operating in higher than the advertised dimension implies state leakage at some point during the evolution (i.e. a trace less than one on the computational space~\cite{WoodGambetta2018}). Throughout the paper we shall use terms state leakage and leakage interchangeably, since our method does not necessarily expose the origin of the state leakage. We cannot distinguish, for example, so-called gate leakage and/or seepage (where population crosses from the computational space to another space~\cite{WoodGambetta2018}) from observables that are influenced by non-computational population that evolves entirely independently.
Our application of the presented method does not quantify the magnitude of the leakage or its contribution to the overall error of the computation: rather, it constitutes a simple and qualitative test of the quantum machine from the user's perspective, complementary to the hardware-specific approaches.

In this paper we refine WP-G's method: equipping it with a robust procedure for dealing with shot noise. We apply this refined method to the IBM Quantum Experience (IBMQ) transmonic quantum computer \textit{ibmqx4} \cite{IBMQC} remotely and reveal evidence of qubit leakage under certain evolutions with a $p$ value less than $5{\times}10^{-4}$.

Sec.~\ref{sec:theory} lays out the theoretical basis of the method of delays, including a new, transparent proof and various example evolutions. Sec.~\ref{sec:methods} introduces our methods for dealing with shot noise; simulations and experimental results are presented and discussed in Sec.~\ref{sec:results}. We draw our conclusions in Sec.~\ref{sec:conclude}.

\section{Method of delays}
\label{sec:theory}
The method of delays begins by assembling an $N {\times} N$ matrix $V$ with elements 
\begin{equation}
V_{kl} = \langle M(k+l-2)\rangle,
\end{equation}
where $\langle M(t) \rangle$ is the time series of an arbitrarily chosen Hermitian observable $M$:
\begin{equation}\label{Expectation}
\langle M(t) \rangle = \Tr(M\mathcal{E}(t)[\rho_0]).
\end{equation}
Here $\mathcal{E}(t)$ describes the time evolution of the initial state $\rho_0$ to the final state $\rho(t)$, for a discrete set of instants $t\in\{0,1,2,\ldots 2N-2\}$. A bound on the dimension will be inferred directly from $V$, which is known as the matrix of delayed vectors. Let $s_i$ be the singular values of $V$, arranged in descending order so that $s_1\geq s_2\geq s_3\ldots$. Since $V$ is real and symmetric, $s_i = |\lambda_i|$ where $\lambda_i$ are the eigenvalues of $V$, sorted by magnitude. Letting $\bm{1}_0$ be an indicator function for the set of zero valued quantities, we have $\text{rank}(V)=\sum_i^N [1-\bm{1}_0(s_i)]=\sum_i^N[1-\bm{1}_0(\lambda_i)]=N-\sum_i^N\bm{1}_0(\lambda_i)$, which clearly shows that the amount of data collected (the number of instants sampled) sets an upper bound on the rank of $V$ --- and hence (as we shall see) on the dimension that may be witnessed. 

It is important to understand that although the LHS of  (\ref{Expectation}) will be directly measured, the operators appearing in the RHS are, by hypothesis, unknown to the {user}. By delegating a quantum program to a remote server, we may attempt to implement \emph{target} evolutions such as $\mathcal{E}(t)[\rho_0]=U^t\rho_0(U^\dagger)^t$, where $U$ is a unitary operator. However, after being compiled into a set of control instructions, transmitted to the remote location and translated (by a possibly noisy, unknown, or even dishonest procedure) into time-varying control fields, we need no longer take the target evolution at face value since the assessment of dimensionality (or leakage) follows only from the returned measurement results and does not rely on any particular choice of control instructions.

WP-G's method~\cite{WolfPerez-Garcia2009} calls for an evolution $\mathcal{E}(t)$ satisfying both:
\begin{enumerate}
\item Stroboscopic homogeneity and Markovianity, {that is,} $\mathcal{E}(t)=\mathcal{E}^t$, and
\item Trace preservation, {that is,} $\textrm{Tr}[\mathcal{E}(\rho)]=\textrm{Tr}[\rho$].
\end{enumerate}
Assumption 1 can be enforced by concatenating any map $\mathcal{E}$ with itself $t$ times.
We introduce the term `stroboscopic Markovianity' to emphasize that the assumption is weaker than full homogeneity and Markovianity, which would require that $\mathcal{E}(s)=\mathcal{E}^s$ for $s\in[0,\infty)$. That is, homogeneity and Markovianity holding on a finer timescale than that which was used to sample the overall evolution-- a needlessly strong assumption. If the dimension of the system is known, the assumption 1 above may be relaxed and WP-G's technique could instead be used to learn the minimal dimension of the environment~\cite{WolfPerez-Garcia2009}. 

The complete positivity of $\mathcal{E}$, namely that it has a Kraus representation $\mathcal{E}(\rho)=\sum_i K_i \rho K_i^\dagger$, is not required as an assumption. This means that not completely positive maps in quantum theory~\cite{ShajiSudarshan2005} or even dynamics beyond quantum theory~\cite{Barrett2007} could be incorporated into the method. More straightforward is the concept of a trace-non-preserving map, intimately related to leakage. In our alternative proof of WP-G's bound on the dimension, the trace preservation assumption will be unnecessary. 

We may rewrite equation \eqref{Expectation} as $\langle M(t)\rangle = \llangle M | T_\mathcal{E}^t | \rho\rrangle$, where $|M\rrangle$ and $|\rho\rrangle$ are elements of a $d^2$-dimensional complex vector space, with inner product $\llangle A|B\rrangle=\textrm{Tr}(A^\dagger B)$~\cite{WolfPerez-Garcia2009}. A bound on the dimension follows quickly from the Cayley-Hamilton theorem. This theorem states that $T_\mathcal{E}$ satisfies its own characteristic equation~\cite{HornJohnson1990}, and implies that the $d^2$th power of $T_\mathcal{E}$ may be written as a linear combination of the $d^2$ lower matrix powers (down to and including the zeroth power):
\begin{align}
T_\mathcal{E}^{d^2}&=\sum_{j=0}^{d^2-1} c_j T_\mathcal{E}^{j}.\end{align}
Multiplying this equation by $T_\mathcal{E}$, it is clear that 
\begin{align}
T_\mathcal{E}^{d^2+1}&=\sum_{j=1}^{d^2} c_j T_\mathcal{E}^{j},
\end{align}
and that by recursive application of this formula, any higher power of $T_\mathcal{E}$ will have a similar decomposition (albeit with different $c_j$). 
We then have $\forall \, q\geq d^2+1$,
\begin{align}
V_{ql} &= \llangle M | T_\mathcal{E}^{q+l-2}| \rho \rrangle = \llangle M | T_\mathcal{E}^{q}T_\mathcal{E}^{l-2}| \rho \rrangle\\
&=\sum_{j=1}^{d^2}c_j\llangle M |  T_\mathcal{E}^{j}T_\mathcal{E}^{l-2}| \rho \rrangle= \sum_{j=1}^{d^2}c_j V_{jl},
\end{align}
and therefore, since all rows with index greater than $(d^2+1)$ are in the space spanned by the first $d^2$ rows, 
\begin{align}
\textrm{rank}(V)\leq d^2.
\label{main_bound}
\end{align}
This enables WP-G's method to certify that a certain evolution $\langle M(t)\rangle$ cannot be described by a quantum system with dimensionality less than $\sqrt{\mathrm{rank}(V)}$, where $V$ is constructed from experimentally gathered data. As with all DWs, some leakage may \emph{not} be witnessed: since the bound~\eqref{main_bound} depends only on the dynamics (represented by $T_{\mathcal{E}}$), and \emph{not} on just a single output state $\rho(t)$, the latter could exhibit state leakage which would go undetected unless it is accompanied by dynamics of the non-computational space that is exposed by the measurement.

Although it is possible to reason about tighter bounds via the minimal polynomial of $T_\mathcal{E}$~\cite{WolfPerez-Garcia2009}, for this purpose we shall find it more instructive to investigate the dense subset of non-defective evolutions.

 \begin{figure*}[]
 \includegraphics[width=\linewidth]{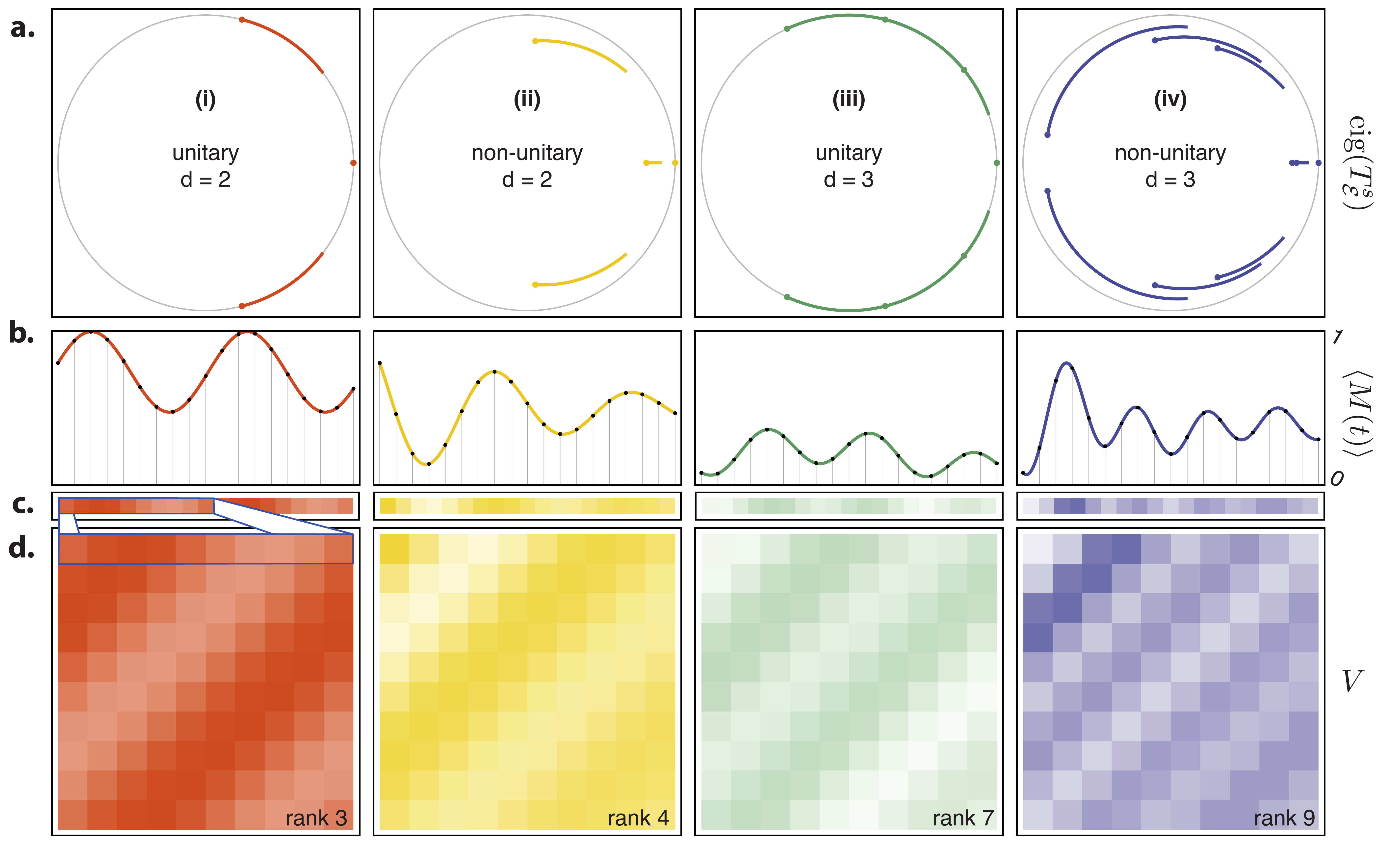}
 \caption{The spectra of nondefective quantum channels determines the maximum complexity of any time series that they generate, and the rank of the corresponding matrix of delayed vectors. Various homogenous, Markovian, completely-positive and trace preserving processes $\mathcal{E}(t)$ are illustrated. (a) The flow of the spectrum $\{\lambda_j^s\}$ of $T^s_{\mathcal{E}}$ is shown in the complex plane for $1\leq s \leq 2$ for  a $d=2$ (i) unitary, (ii) non-unitary and $d=3$ (iii) unitary and (iv) non-unitary process. Where $|\lambda_j|=1$, the flow is on the unit circle; where $|\lambda_j|<1$ there is an inexorable spiral towards the origin. All of these processes feature a stationary eigenvalue at 1 because of their trace preserving nature. The time varying expectation $\langle M(t)\rangle$, defined by a rank one initial state $\rho$, rank one measurement operator $M$ and $T_{\mathcal{E}}^t$, is shown in (b) and (c) for discrete instants $t=0,1,2\ldots 18$. Smooth curves show $\langle M(t)\rangle$ for continuous $0\leq s\leq 18$, and are a guide for the eye. Oscillation frequencies, phase offsets, decay envelopes and vertical shifts may be traced back to the spectral properties of $T_{\mathcal{E}}$. Indeed, the different frequencies were also noted in the time series of classical dynamical systems~\cite{Packard1980}. The matrix of delayed vectors $V$, whose rank (in these non-defective cases) is no more than the number of distinct eigenvalues of $T_{\mathcal{E}}$, is shown in (d). If  $U$ is a matrix drawn from the Haar ensemble and $Y=U^{1/3}$, unitary evolutions were constructed as $T_{\mathcal{E}}=Y\otimes \bar{Y} $ and non-unitary evolutions as $T_{\mathcal{E}}=0.08T_{B}+0.92 (Y\otimes \bar{Y} )$; $T_{B}$ is a random CPTP map generated via the algorithm in Ref.~\cite{BruzdaCappelliniSommers2009}. \label{matrix_plot}}
 \end{figure*}

\subsection{Specialisation to non-defective evolutions}
Assume the superoperator $T_{\mathcal{E}}$ may be spectrally decomposed as $T_{\mathcal{E}}=\sum_j \lambda_j |\lambda_j\rrangle \llangle \lambda_j| $ with complex eigenvalues $\lambda_k$ and orthonormal eigenvectors $\llangle \lambda_i|\lambda_j\rrangle=\delta_{ij}$.  This leads to 
\begin{equation}
\langle M (t)\rangle  = \sum_{j=1}^{d^2} \lambda_j^t \llangle M |\lambda_j\rrangle\llangle \lambda_j |\rho\rrangle.
\label{linear_combination}
\end{equation}
We have assumed here that $T_{\mathcal{E}}$ is normal, or diagonalizable, i.e not defective. Almost all matrices are not defective, with each eigenvalue having algebraic multiplicity equal to geometric multiplicity. 

The matrix of delayed vectors $V$ now has elements
\begin{align}
V_{kl}& = \sum_{j=1}^{d^2}  \llangle M |\lambda_j\rrangle\llangle \lambda_j |\rho\rrangle\lambda_j^{k+l-2}
 =: \sum_{j=1}^{d^2}V_{kl}^{(j)}.
 \label{d_sq_terms}
\end{align}
An alternative route to the bound on the rank of $V$ may be found by bounding the rank of the individual matrices $V^{(j)}$. The latter are easily seen to be rank 1:
\begin{align}
V_{ml}^{(j)}&= \llangle M |\lambda_j\rrangle\llangle \lambda_j |\rho\rrangle\lambda_j^{m+l-2}=\lambda_j^{m-1} \llangle M |\lambda_j\rrangle\llangle \lambda_j |\rho\rrangle\lambda_j^{l-1} \nonumber
\\&= \lambda_j^{m-1} V^{(j)}_{1l},
\end{align}
since all rows are proportional to the first row. Because of the subadditivity of matrix rank and the number of terms in the sum in {Eq.~\eqref{d_sq_terms}}, we recover
$
\mathrm{rank}(V)\leq d^2.
$

It is also clear that the sum in Eq.~\eqref{d_sq_terms} need only run over the distinct eigenvalues of $T_{\mathcal{E}}$, such that a tighter upper bound on the rank of $V$ may be found. For example, assume $\lambda_k=\lambda_j, k\neq j$. Then $W^{(j)}:=V^{(j)}+V^{(k)}$ is a matrix with elements
\begin{align}
W^{(j)}_{ml} &= \llangle M |\lambda_j\rrangle\llangle \lambda_j |\rho\rrangle\lambda_j^{m+l-2} + \llangle M |\lambda_k\rrangle\llangle \lambda_k |\rho\rrangle\lambda_k^{m+l-2}  \nonumber\\
& = \llangle M |\lambda_j\rrangle\llangle \lambda_j |\rho\rrangle\lambda_j^{m+l-2} + \llangle M |\lambda_k\rrangle\llangle \lambda_k |\rho\rrangle\lambda_j^{m+l-2} \nonumber\\
& = \left[\llangle M |\lambda_j\rrangle\llangle \lambda_j |\rho\rrangle+ \llangle M |\lambda_k\rrangle\llangle \lambda_k |\rho\rrangle\right]\lambda_j^{m+l-2}\nonumber\\
& = \lambda_j^{m-1}W_{1l}^{(j)};
\end{align}
i.e. it is also rank one but replaces two matrices in the sum in Eq.~\eqref{d_sq_terms}.

Specific evolutions may have fewer than the maximal number of distinct eigenvalues. Trace preserving maps satisfy $\llangle \mathbb{I}|T_{\mathcal{E}}|\rho\rrangle=\llangle \mathbb{I}|\rho\rrangle$ which implies $\llangle \mathbb{I}|T_{\mathcal{E}}=\llangle\mathbb{I}|$.  Then $T_{\mathcal{E}}$ has at least one eigenvalue equal to $1$, and the corresponding matrix $V^{(j)}$ would therefore be a constant `zero frequency, unit magnitude' matrix.  This need not impact on the rank of $V$: if there are $x$ linearly independent eigenvectors with eigenvalue 1, however, then $x$ of the $V^{(i)}$ are aggregated into the zero frequency matrix.

Fig.~\ref{matrix_plot} provides a visualisation of typical (unitary and non-unitary) evolutions in $d=2,3$. Also shown are the corresponding delayed-vector matrices and the flow of the eigenvalues $\{\lambda_j^s\}$ of $T_\mathcal{E}^s$. 

\subsection{Specialisation to unitary evolutions}
\label{unitary_special}
Unitary evolutions only have a single Kraus operator in their operator-sum representation, so that a matrix representation exists where $T_{\mathcal{E}}=\sum_i K_i\otimes\bar{K_i}=U\otimes \bar{U}$~\cite{WolfPerez-Garcia2009,Wolf2012}. $\bar{U}$ denotes the complex conjugate of $U$. It is easy to see that the eigenvalues of $T_{\mathcal{E}}$ are $e^{i[\omega_l-\omega_k]}, \quad l,k=1\ldots d$, with eigenvectors $|lk\rrangle$, where $e^{i\omega_l}$ are the eigenvalues of $U$. It then immediately follows that $d$ eigenvalues (when $l=k$) are equal to $1$, and therefore 
\begin{align}
\mathrm{rank}(V) \leq d^2-d+1.
\label{tighter_bound_unitary}
\end{align}
To see this, note that
\begin{align}
\langle M (t)\rangle  &= \sum_{l,k} e^{it[\omega_l-\omega_k]} \llangle M |lk\rrangle\llangle lk|\rho\rrangle \nonumber \\
&=\sum_{m=1}^d\llangle M |mm\rrangle\llangle mm|\rho\rrangle \nonumber\\
&\phantom{=}+ \sum_{l\neq k} e^{it[\omega_l-\omega_k]} \llangle M |lk\rrangle\llangle lk|\rho\rrangle.
\end{align}
Since $T_{\mathcal{E}}$ is unitary it is also non defective. The eigenvalue 1 has multiplicity at least $d$, and $|mm\rrangle$ are a linearly independent set of vectors spanning the corresponding eigenspace. 

A time independent, zero frequency term absorbs some of the degrees of freedom of $T_{\mathcal{E}}$, leaving up to $d^2-d$ further terms with unique `beating' angular frequencies, with each positive frequency matched with a negative one~\cite{Wolf2012,BengtssonZyczkowski2006}.

 The rank may also be reduced by certain choices of $M,\rho$ and $U$. If either $M$ or $\rho$ commute with $U$, then the second sum vanishes since  e.g. $\llangle M|lk\rrangle=0$: the evolution is then constant and the rank of $V$ becomes 1. 

Similarly, if $U$ is degenerate (i.e. $\omega_l=\omega_k, l>k$),  degrees of freedom will be lost: the maximum rank of $V$ is therefore $2{\times}\binom{q}{2}+1$ -- where $q$ is the number of distinct eigenvalues of $U$. In such a situation, $U$ is an embedding of SU$(q)$ into SU$(d)$: the most extreme case being when $q=1$ and $U\rightarrow\mathbb{I}$ (an embedding of U$(1)$) . 

One method for ensuring the eigenvalues of a target unitary matrix are distinct is to draw them from the Haar (circular unitary) ensemble. Single qubit Haar-random unitary operators may be constructed as follows: 
\begin{equation}\label{randommatrix}
U = \left(
\begin{array}{cc}
\cos\phi e^{i\psi} & \sin\phi e^{i\chi} \\
- \sin\phi e^{-i\chi} & \cos\phi e^{-i\psi}\\
\end{array}
\right),
\end{equation}
where the Euler angles $\psi$ and $\chi$ are chosen uniformly in the intervals $0 \leq \psi < 2\pi$, $0 \leq \chi < 2\pi$, and $\phi = \sin^{-1}(\xi^{1/2})$ with $\xi$ chosen uniformly in the interval $0 \leq \xi < 1$. The generalisation to arbitrary dimension may be found in Ref.~\cite{ZyczkowskiKus1994}, and involves $d^2$ angles. Such a technique ensures with high probability that no more than $d$ of the eigenvalues of $T_{\mathcal{E}}$ are equal to 1.

While the preceding arguments allow one to predict the rank of $V$ from a known evolution, it is of course our aim to learn something of the dimension of the evolution from an experimentally measured $V$ without assuming unitarity. We therefore proceed with the minimal assumptions (namely stroboscopic homogeneity and Markovianity), and rely on Eq.~\eqref{main_bound} to draw conclusions about $d$.

 \subsection{Irreducible dimension witnesses}
One may ask whether the method of delays is an \emph{irreducible} dimension witness~\cite{CongCaiBancal2017} of a quantum system, namely, one that allows processes admitting various tensor product decompositions to be mutually distinguished. For instance, processes of the form $\mathcal{E}_1\otimes\mathcal{E}_2$ from $\mathcal{E}$ (those without a product structure) but with $\textrm{dim}T_{\mathcal{E}_1\otimes\mathcal{E}_2}=\textrm{dim}T_{\mathcal{E}}$~\footnote{DWs that allow all such possible decompositions to be distinguished are termed~\emph{gamut} dimension witnesses~\cite{AguilarFarkasMartinez2018}.}. Removing the controlled-NOT gates from Fig.~\ref{dimCircuit}b will convert it from the latter class to the former. We show in Appendix~\ref{not_irreducible_dw} that in general, the method of delays is not an irreducible DW. When the states, measurements, and evolutions all share the same product structure, the rank of the matrix of delayed vectors may still reach up to $d^2$. When the assumption of unitary evolutions hold, the situation changes and the method becomes an irreducible DW.

\section{Dealing with shot noise}
\label{sec:methods}
Experimental noise imposes some important considerations on the method of delayed vectors. The original work on the method of delays already noted the paramount importance of low-noise data in determining the dimensions of classical chaotic systems~\cite{Packard1980}. 
 In a quantum computer, noise arises from numerous sources including decoherence, finite temperature effects, undesirable crosstalk between components, and limited user runtime. Many of these do not affect the method of delayed vectors because they do not violate the assumption of stroboscopic homogeneity and Markovianity. Most importantly, however, finite statistics imply (for nontrivial evolutions) a full-rank matrix of delayed vectors with high probability, meaning that some method of validation (of the `true' singular values of that matrix) or discreditation (of the noisy ones) is necessary for the method to be effective. Erroneous signals arising from finite statistics are unavoidable and blight all DWs, and can mask any small signatures of leakage.
One cannot therefore completely avoid the risk of false positives: low dimensional systems classified as being much higher dimensional simply due to unavoidable statistical noise in their observable dynamics. The validation method we shortly introduce offers to redress the situation, but at the increased risk of false negatives. 

WP-G suggest that a validation threshold may be set by the estimated amount of errors~\cite{WolfPerez-Garcia2009}. We complete this suggestion in a way that preserves model independence, by taking a worst case view of statistical errors. Following WP-G, take the observed data after $n$ shots $M'=M+M_{\epsilon}$ for some perturbation $M_{\epsilon}$. $V'$ and $V_{\epsilon}$ (with respective singular values $s_i'$ and $s_i^{(\epsilon)}$ once more ordered by decreasing magnitude) are defined analogously. We may bound the singular values of $V_{\epsilon}$~\cite{Garren1968}
\begin{align}
s^{(\epsilon)}_i &\leq N \max_t |\langle M_{\epsilon}(t)\rangle|.
\end{align}
Next, assume (for simplicity) that $M$ has eigenvalues $\pm1$ with associated success probabilities of $p_\pm(t)$. The experimentally measured value $|\langle M'(t)\rangle|=|(n_+-n_-)/n|=|2n_+/n-1|$ where $n_\pm$ are the number of successes associated to each eigenvalue. Now the distribution of $n_+$ is binomial with mean $np_+$ and variance $np_+(1-p_+)$, implying that $\langle M'(t)\rangle$ has mean $2p_+-1$ and variance $4p_+(1-p_+)/n$. All of this implies that $N\max_t\langle M_{\epsilon}(t)\rangle$ has mean zero and variance $\sigma^2=4N^2p_+(1-p_+)/n$. Setting a tolerance threshold 
$
h:=zN/ \sqrt{n},
$
we have
\begin{align}
\textrm{Pr}(s^{(\epsilon)}_i \geq h)&\lessapprox 1-\textrm{erf}\left(\frac{h}{\sigma\sqrt{2}}\right)\\
&= 1- \textrm{erf}\left(\frac{h\sqrt{n}}{2N\sqrt{2p_+(1-p_+)}}\right)\\
&\leq 1- \textrm{erf}\left(\frac{h\sqrt{n}}{N\sqrt{2}}\right)\\
&\leq 1- \textrm{erf}\left(\frac{z}{\sqrt{2}}\right).
\label{1merf}
\end{align}
In the first line, we used a two-tailed test for a normal approximation to the binomial distribution. In the third line, we took the worst case $p_+{\to}1/2$~\footnote{This actually improves the normal approximation, which is more accurate when $np_+\gg 1$ and $n(1-p_+)\gg1$.}, meaning that the final expression is thus a model-independent, upper bound on the $p$ value. Here $z$  is the number of standard deviations, and controls the desired significance level. 
 Weyl's inequalities~\cite{Bhatia1999} yield
 \begin{align}
 s_i&\geq s'_{i+j-1} - s^{(\epsilon)}_j\quad \forall j\\
 &\geq s'_{i} - s^{(\epsilon)}_1.
 \label{weyl}
 \end{align}
 Combining~\eqref{1merf} and \eqref{weyl} and replacing $s^{(\epsilon)}_1$ by the threshold $h$ (which is an upper bound with high probability) we have 
 \begin{align}
 \textrm{Pr}(s_i \leq \max(0,s_i'-h)) \leq 1- \textrm{erf}\left(\frac{z}{\sqrt{2}}\right).
 \end{align}
 This leads us to an estimate of the true rank 
\begin{align}
\textrm{validatedRank}(V',z)&:=\min\{k|s'_{k+1}\leq N z /\sqrt{n}\}.
\label{stat_threshold}
\end{align}
We used the fact that singular values are always positive. The probability of the true rank being lower than this estimate is correspondingly lowered as $z$ increases.
Note how the threshold scales with the size of the delayed-vector matrix, but exhibits a $1/\sqrt{n}$ scaling with respect to the number of shots. After fixing a desired $p$ value, the threshold is then set. We set a $p$ value level of 0.001 (corresponding to $z=3.29$). In decision-theoretic language, this sets a significance level at which to reject the null hypothesis. The null hypothesis is that the device
has the advertised quantum dimension $d_a$, and evolves in a time-homogenous, Markovian manner. We calculate a conservative upper bound on the $p$ value for the null hypothesis by assuming the worst case shot noise. The worst-case shot noise is realised when e.g. $\langle M_\epsilon(t)\rangle=1 \quad \forall t$. We emphasise that our approach gives a conservative, upper bound on the $p$-value, and there is some scope for tighter bounds that increase the statistical power of the test. (The statistical power of a binary hypothesis test is the probability of rejecting the null hypothesis -- e.g. that $d=d_a$ --  when it is false~\cite{everitt1995cambridge}). Nevertheless, our approach will tend to underestimate the significance of the experimental findings. A further consideration is the dependence on $N$, which implies that the matrix of delayed vectors should be chosen large enough to test the desired dimension boundary, but no larger (since that raises the tolerance and increases the false negative rate). This will become an issue later when we apply the dimension witness on a remote quantum processor.

It is also possible to let the data itself dictate a tighter bound on the $p$ value. Let $d_a$ be the advertised dimension. The probability of a noisy singular value reaching $s'_{1+d_a^2}$ or more is the $p$ value for rejecting the null hypothesis:
\begin{align}\label{pvalue}
\textrm{Pr}\left(s^{(\epsilon)}_i\geq s'_{1+d_a^2}\right)=1-\textrm{erf}\left(\frac{\sqrt{n}s'_{1+d_a^2}}{N\sqrt{2}}\right).
\end{align}


\begin{figure}[]
\includegraphics[width=\columnwidth]{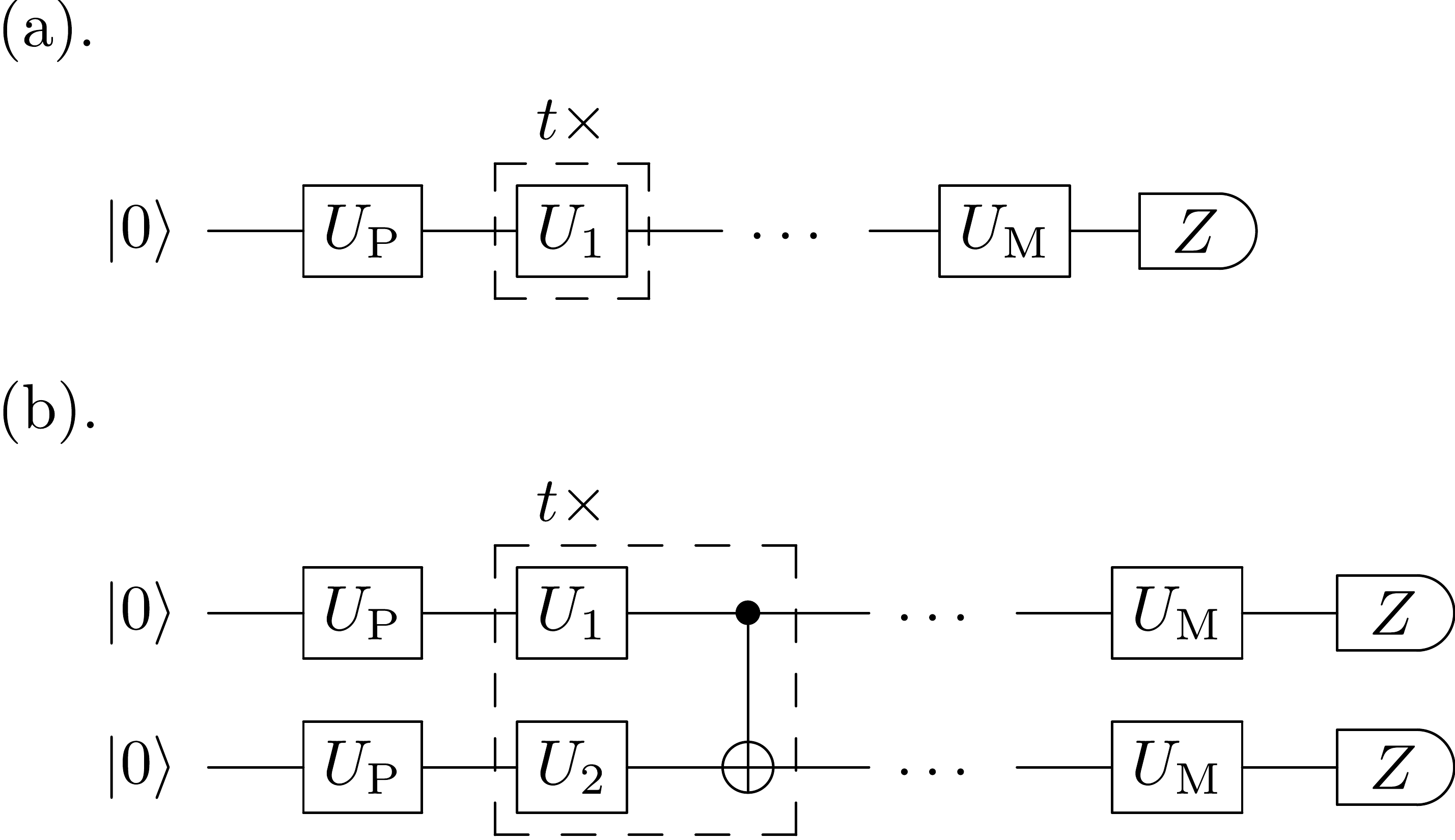}
\caption{\label{dimCircuit} Circuits representing target (a) single qubit and (b) two qubit unitary evolutions. In each case the highlighted region shows a single step that is repeated $t$ times, where $t\in\{0,1,\ldots 2N-2\}$. Single-qubit gates are defined by three angles, and the two-qubit gate is a controlled-NOT. }
\end{figure}

\section{Experimental results from IBMQ}
\label{sec:results}
We now apply the method of delays, along with our refinements, on the publicly accessible IBMQ quantum computer \textit{ibmqx4} \cite{IBMQC}.
The components of this device are transmons -- weakly anharmonic, resonant superconducting circuits exhibiting a countable but potentially infinite number of energy states~\cite{KochYuGambetta2007}. Their anharmonicity, derived from non-linear circuit elements called Josephson junctions, implies that transitions into unwanted energy levels are detuned from the targeted microwave control frequencies. Ideally such unwanted transitions never occur: but spectral congestion, weak non-linearity and finite temperatures imply that they will with potentially non-negligible probability. Experimental techniques have been developed to minimise leakage~\cite{MotzoiGambettaRebentrost2009,McKayWoodSheldon2017}. Leakage may be characterised in a model-specific way by engineers utilising a direct measurement of non-computational states~\cite{McKayWoodSheldon2017, Chen2016}, but this operation is not available to remote users.

In this device, all qubits start in their fiducial, computational and ground state $\rho=|0\rangle\langle 0 |$. The following, single-qubit target unitary operations are drawn from the Haar ensemble: $U_{\mathrm{P}}$ selects the initial state $\rho_0$, $U_{\mathrm{M}}$ selects the measurement $M$ and $U_i, \, i=1,2$ describe single qubit gates. For single qubit experiments, $\mathcal{E}[\rho] = U_1\rho U_1^\dagger$ and for two qubit experiments $\mathcal{E}[\rho\otimes\rho] = C_{X}^{(1,2)}(U_1\otimes U_2)(\rho\otimes \rho)(U_1\otimes U_2)^\dagger C_{X}^{(1,2)},$ where $C_{X}^{(1,2)}$ is a controlled-NOT gate~\cite{NielsenChuang2004} controlled on qubit $1$ and targeted on qubit $2$. $\mathcal{E}$ is repeated $t$ times as shown in Fig. \ref{dimCircuit} before being measured in the computational $Z$ basis. 

\begin{figure}[]
	\includegraphics[trim={0cm 0cm 0cm 0cm },clip,width=0.5\textwidth,height=\textheight, keepaspectratio]{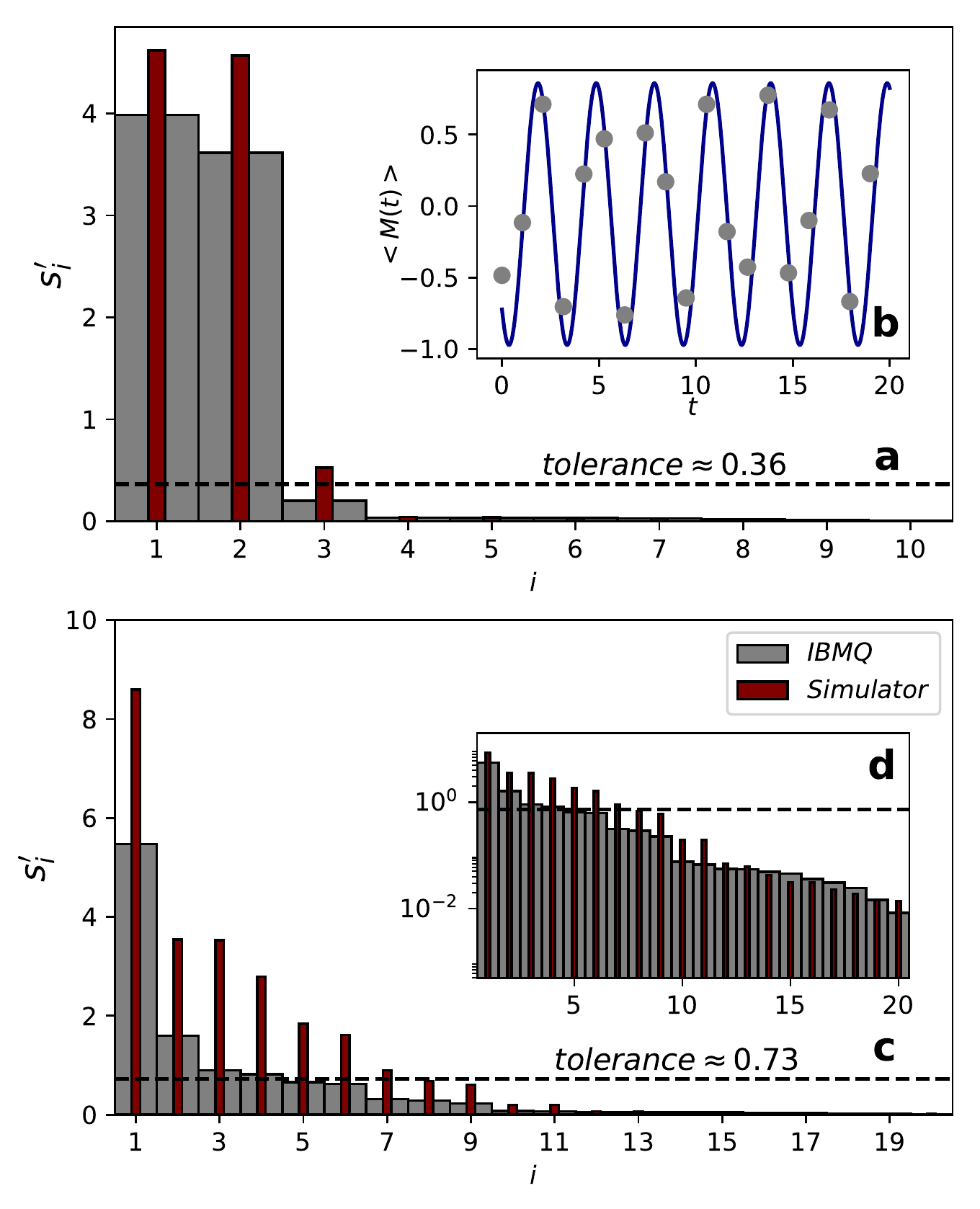}
	\caption{The observed singular values for (\textbf{a}) single qubit and (\textbf{c}) for two qubit evolutions (shown in logarithmic scale for ease of comparison in the inset (\textbf{d})). The grey dots in inset (\textbf{b}) show the time series of an example evolution for a single qubit, with the continuous blue line showing a continuous time simulation of the target evolution.
	Thick grey bars correspond to the IBMQ device, while thin red bars correspond to a local simulation of the target evolution. The dashed line shows the tolerance level at a $p$ value of 0.001 (see Eq.~\eqref{stat_threshold} ).}
	\label{1DimFig}
\end{figure}

Target circuits were compiled into the openQASM language~\cite{CrossBishopSmolin2017}, and submitted for execution on IBM's device remotely. For the single qubit experiments, control instructions known as barriers were inserted between the $U_1$ gates to prevent IBM's compiler from simplifying the circuit. Assuming these barrier commands are obeyed, we expect the assumption of stroboscopic homogeneity and Markovianity to be valid in this system, since the instructions for each of the repeated target evolutions are identical.  Arbitrary single qubit gates (native to each qubit) are specified by three angles and controlled-NOT gates by specifying control and target qubits (native, up to a limited chip topology). An example of remotely accessed device data, along with a locally-run simulation, is shown in Fig.~\ref{1DimFig}. 

\begin{figure*}\label{ExpData}
	\includegraphics[trim={0cm 0cm 0cm 0cm },clip,width=\textwidth,height=\textheight, keepaspectratio]{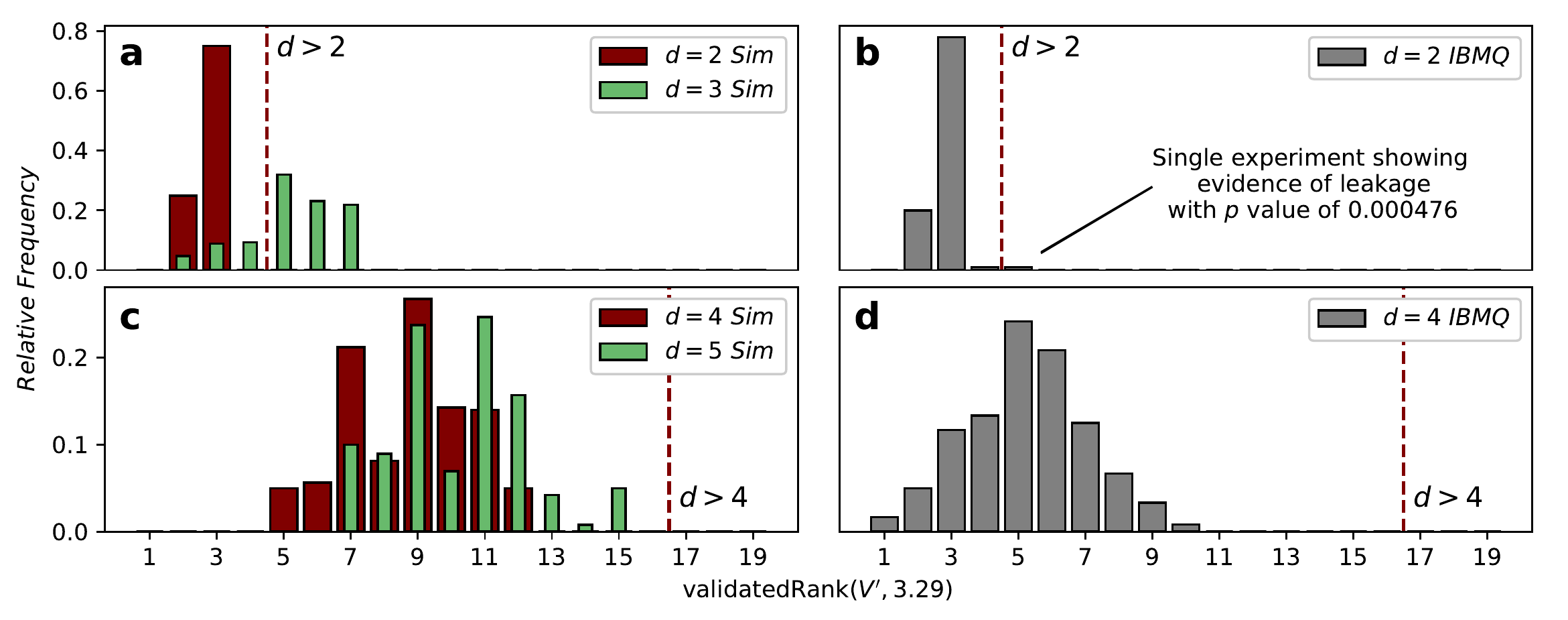}
	\caption{Number of validated singular values from a single qubit (\textbf{a}) 8192-shot ideal simulation (red bars) and (\textbf{b}) 8192-shot execution on IBMQ (grey bars). Two qubit simulations and IBMQ data are shown in  (\textbf{c}) and (\textbf{d}) respectively. {Thin light green bars show 8192-shot ideal simulations of random unitary experiments in a (\textbf{a}) 3 dimensional and (\textbf{c}) 5 dimensional system. }The thresholds from Eq.~\eqref{stat_threshold} were set at a $p$ value of $0.001$. The dashed lines show an upper bound on the rank of a delayed vector matrix generated by a $d = 2$ (in \textbf{a}, \textbf{b}) or $d = 4$  (in \textbf{c}, \textbf{d}) dimensional system.}
	\label{histograms}
	\centering
\end{figure*}

Firstly, we locally simulated 2000, 8192-shot (but otherwise idealised) experiments for random unitary evolutions for $d=2,3$ (see  Fig. \ref{histograms}\textbf{a}), and $d=4,5$ (see Fig. \ref{histograms}\textbf{c}) that mimic the target evolution to be applied to transmons. We applied our  validation condition Eq.~\eqref{stat_threshold}, and the distribution over validated singular values represents an expectation for ideal experiments of the corresponding dimension and number of shots. Since the transmons are expected to be evolving as approximate qubits, the behaviour of the real device under the target evolutions described in circuits in Fig.~\ref{dimCircuit} should match $d=2,4$ (respectively) more closely than $d=3,5$.


\subsection{Single qubits}
Fig. \ref{histograms}\textbf{a} indicates that the statistical power of our test, using 8192 shots, is sufficiently high to provide a good chance of exposing single-qubit leakage if it exists. This is because in over half of the simulated $d=3$ experiments, the null hypothesis (that $d=2$) is rejected with $p$ value no more than our chosen value of $0.001$.

Twenty experiments, as depicted in Fig.~\ref{dimCircuit}a, of 8192 shots each were performed remotely during {2018.10.26 to 2018.10.28} for a randomly selected initial state (via $U_{\mathrm{P}}$), measurement (via $U_{\mathrm{M}}$) and evolution (via $U_{1}$) for each qubit of the IBMQ 5-qubit device \textit{ibmqx4} version 1.3.0~\cite{IBMQC}. We constructed a $10{\times}10$ matrix $V$, and applied our refined WP-G analysis. The results are shown in Fig.~\ref{histograms}\textbf{b} and show evidence of leakage in a single case at a $p$ value no greater than 0.001.

Re-examining this single experiment, we calculate a tighter upper bound on the $p$ value using Eq.~\eqref{pvalue}, and find it to be 0.000476. This is significant evidence that IBM's transmonic  `qubits' evolve in a space that is higher dimensional than advertised. Note that the small $p$ value represents a large statistical significance, but this should not be confused with the magnitude of the effect itself.

\subsection{Two qubits}
To explore the possible leakage from higher dimensional computational spaces, we next target two-qubit evolutions. Two qubit gates are more challenging experimentally, usually exhibiting far worse errors compared to single qubit gates~\cite{Ballance2016}, and potentially suffering more from leakage~\cite{McKayWoodSheldon2017}. 
In a similar fashion to the single qubit experiments we constructed a circuit from 2 random unitary single qubit gates (acting on each qubit) interleaved with a controlled-NOT gate to target a partially entangling evolution, as depicted in Fig.~\ref{dimCircuit}b. As above, the initial state and measurement were randomised by selecting  random single qubit gates $U_{\textrm{P}}$ and $U_{\textrm{M}}$. 

A pseudo-random two-qubit unitary may be implemented by repeating the gate sequence enclosed in dashes in Fig.~\ref{dimCircuit}b. The final unitary approaches the Haar measure exponentially in the number of repetitions~\cite{EmersonLivineLloyd2005}, and it may take only tens of repetitions to attain a good approximation for a two-qubit unitary~\cite{EmersonWeinsteinSaraceno2003}. The decoherence and relaxation times of the IBMQ transmon qubits limit us to a single repetition per time step $t$. The resulting two-qubit unitaries we implement are thus unlikely to be pseudo-random. We have numerically sampled one million two-qubit unitaries using the prescription in the highlighted area in Fig.~\ref{dimCircuit}b and in all instances found $|\lambda_i-\lambda_j|>1{\times}10^{-15}\quad \forall i>j$, where $\lambda_i$ are the eigenvalues of the sampled matrix. This makes the chosen unitaries `optimal' in the sense of saturating Eq.~\eqref{tighter_bound_unitary}.

Twenty experiments, as depicted in Fig.~\ref{dimCircuit}b, of 8192 shots each were performed remotely during {2018.10.28 to 2018.10.31} on each qubit pair whose connectivity via controlled-NOT is allowed on the IBMQ 5-qubit device \textit{ibmqx4} version 1.3.0~\cite{IBMQC}. We chose to construct a $20{\times}20$ matrix $V$ to keep $N$ low while avoiding clipping the maximum witness-able dimension.

To justify the choice of a low value for $N$, consider our treatment of shot noise. Because of the worst-case analysis, the tolerance level is proportional to $N$. The singular values of experimental time series, however, will not generally be the worst case and will be subject to a lower upper bound. The tolerance level upper bound will become looser as $N$ increases, which could lead to needless reduction in statistical power.

Fig. \ref{histograms}\textbf{c} presents the result of our 8192-shot simulations, and shows the impossibility of distinguishing whether the advertised two-qubit system was four- or five-dimensional. Indeed, we found no evidence of leakage (or reason to reject the null hypothesis) in the data shown in Fig.~\ref{histograms}\textbf{d}. 

This inconclusive outcome is not unexpected given the low statistical power of the test due to the relatively low number of 8192 shots. In Appendix~\ref{app:B}, we illustrate the increased statistical power obtainable with a higher number of shots, by simulating 4- and 5-dimensional experiments. From this analysis we estimate that at least $2^{18} \sim 3{\times}10^5$ shots are needed to allow the test to have sufficient power to investigate the $d>4$ borderline. Given current limitations of the IBMQ interface, this is highly time consuming due to the overhead time of submitting, compiling, queuing and retrieving jobs. 

\section{Discussion}
\label{sec:conclude}

We have developed the method of delayed vectors for quantum systems and applied it to test leakage in the transmon qubits of the remotely-accessed IBMQ device. 
We infer evidence of leakage from our single-qubit experiments with high statistical confidence. Remarkably, we relied only on a limited, public-facing interface to a remote quantum computer to reach our conclusions: using nothing other than (what are advertised as) single-qubit gates and measurements. For our two-qubit experiments, the lack of sufficient statistical power prevents us from conclusively establishing the absence or presence of leakage. This could be remedied in future experiments. 

In the emerging era of remotely-accessed quantum computing devices, the principal advantage of the method of delays lies in its model and device independence. The task of witnessing the dimension using this method can be readily automated for users without expertise in specific quantum hardware platforms. The user's ability to place a rigorous upper bound on the $p$ value for rejecting the null hypothesis that the dimension is as advertised should be welcome.

Compared to other DWs, the appeal of the method of delays lies in the lack of a user's need to trust the components of the quantum device. The required assumptions of stroboscopic homogeneity and Markovianity can be independently tested~\cite{PollockRodriguez-RosarioFrauenheim2018} to further increase the confidence in findings such as those we have presented. Similar to our treatment of shot noise, spurious singular values arising from finite levels of non-Markovianity could be bounded and discarded if sufficiently small. 
Compared to randomised-benchmarking-based approaches to leakage detection, the main drawback of the method of delays lies is the need for a large circuit depth. Indeed Eq.~\eqref{main_bound} implies that a circuit depth of at least $2d_a^2$ is required to detect leakage beyond an advertised dimension of $d_a.$

The method of delays may become more effective at exposing leakage by implementing evolutions $T_{\mathcal{E}}$ with distinct eigenvalues, each having the same weight in $\langle M(t)\rangle$. Given that Eq.~\eqref{main_bound} is a greater lower bound than~\eqref{tighter_bound_unitary}, non-unitary processes can introduce more singular values into $V$ than unitary ones, and therefore may be a more optimal target for exposing leakage. 

However, the eigenvalues of a non-unitary $T_{\mathcal{E}}$ will tend to be closer together, which reduces the robustness to noise by bringing the evolution closer to a lower dimensional one. In fact, the most informative evolutions would likely arise from modelling the physics of the system in question plus its control apparatus -- for example driving transitions most likely to lead to leakage. 

Our results from the remotely-accessed IBMQ illustrate the limiting role of statistics due to shot noise in a model-independent detection of leakage. Depending on our chosen tolerance level, this leads to false positives and false negatives. As quantum devices improve, the demand for lower thresholds will raise the cost of model-independent detection leakage. Eventually, quantum leakage detection will require formal quantum verification methods analogous to those regularly deployed for classical hardware verification. 

In the mean time, the method of delays should be a useful way to estimate the effective dimensionality of a broad range of quantum dynamical systems. Its model-independence should be useful in comparing different quantum computing hardware platforms such as trapped ions or silicon photonics, as well as helping to determine the minimum complexity required to model various systems in condensed matter physics or quantum biology: where the Hilbert space of a complex quantum system (such as a photosynthetic light harvesting complex) is unclear and often an issue of debate~\cite{LambertChenCheng2013}.

\section{Acknowledgments}
We thank Theodoros Kapourniotis for helpful discussions, Tom Close for pointing out the utility of the Cayley-Hamilton theorem, and Dominic Branford for feedback on this manuscript.
We acknowledge use of the IBM Q for this work. The views expressed are those of the authors and do not reflect the official policy or position of IBM or the IBM Q team. G.C.K was supported by the Royal Commission for the Exhibition of 1851. A.D. was supported by the EPSRC (EP/K04057X/2) and the UK National Quantum Technologies Programme (EP/M013243/1). The python code used to generate openQASM scripts and interface with the IBM API is available upon request from the authors. 
\appendix

\begin{figure}[]
\label{shotsSim}
	\includegraphics[trim={0cm 0cm 0cm 0cm },clip,width=0.5\textwidth,height=\textheight, keepaspectratio]{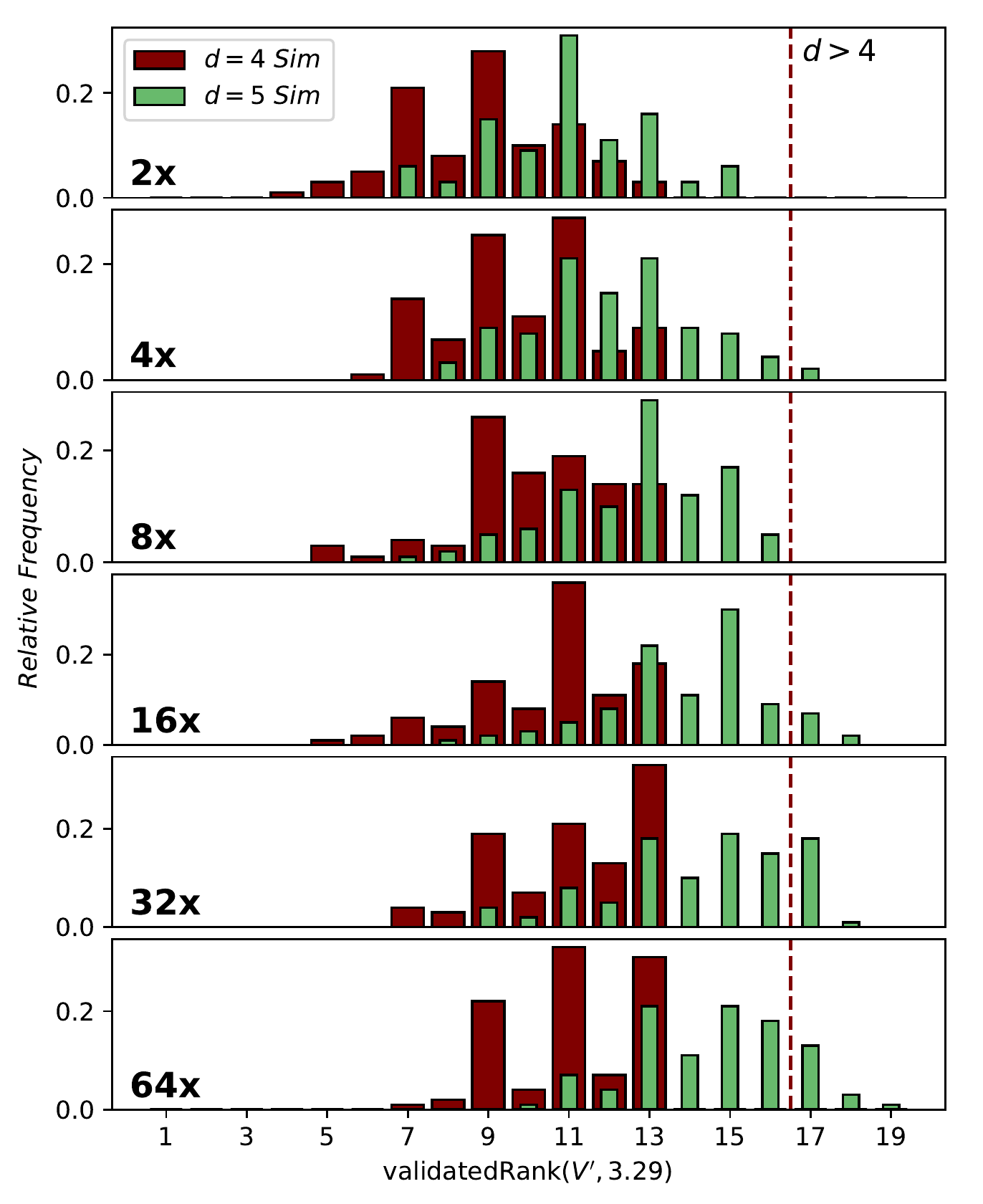}
	\caption{Number of validated singular values via a finite-shot ideal simulation of a 4-dimensional system (thick red bars) and a 5 dimensional system (thin light green bars) for different number of experimental shots. The thresholds from Eq.~\eqref{stat_threshold} were set at a $p$ value of $0.001$. The multiplication factor for the number of shots used, compared to the main experiment which used 8192 shots, is shown in the lower left corner of each panel.}
	\label{shotsSim2}
	\centering
\end{figure}

\section{Singular values from simulated 4- and 5-dimensional systems}
\label{app:B}

One hundred $2^x$ shot ideal simulations with random initial state, measurement operator and evolution were generated with {$x\in \{14,15,16,17,18,19\}$ }for 4- and 5-dimensional systems. The results are shown in Fig. \ref{shotsSim2} and illustrate the increased statistical power of the test as more shots are gathered.

\section{Delayed vector method not an irreducible dimension witness}
\label{not_irreducible_dw}
In the case of an evolution composed of uncorrelated preparations, evolutions and measurements we have
\begin{align}
\langle M(t)\rangle&=\Tr \left(\bigotimes_i M_i\mathcal{E}_i^t[\rho_i]\right)\\
&=\prod_i \llangle M_i |T_{\mathcal{E}_i}^t|\rho_i\rrangle\\
&=\prod_i\sum_{j=1}^{d_i^2} [\lambda_j^{(i)}]^t \llangle M_i |\lambda_j^{(i)}\rrangle\llangle\lambda_j^{(i)}|\rho_i\rrangle\\
&=\sum_{j_m=1}^{d_m^2} \ldots \sum_{j_1=1}^{d_1^2}  \left[\prod_i\lambda_{j_i}^{(i)}\right]^t \prod_i\llangle M_i |\lambda_{j_i}^{(i)}\rrangle\llangle\lambda_{j_i}^{(i)}|\rho_i\rrangle, \nonumber
\end{align}
where $d_i$ is the dimension of the $i$th such component evolution. Now as long as all $\lambda_j^{(i)}$ are unique, and moreover that no $\lambda_j^{(i)}$ is a product of some of the others, then $\prod_i\lambda_{j_i}^{(i)}$ are all unique. Then this expression has the same form as Eq.~\eqref{linear_combination} and is a linear combination of up to $\prod_id_i^2 = (\prod_i d_i)^2$ unique eigenvalues. This is the same maximal number of unique eigenvalues as for an irreducible evolution in the whole space having dimensions $\prod_i d_i$. This is a counterintuitive result since in the independent case much of the total Hilbert space remains unexplored. We sacrificed no loss in generality by choosing non-defective evolutions, since any counterexample is sufficient to rule out an irreducible witness.

The picture is subtly different with unitary evolutions. Re-applying the arguments of Sec~\ref{unitary_special}, we see from the above expression that $\textrm{rank}(V)\leq \prod_i(d_i^2-d_i+1)$. This is a a tighter upper bound than $\textrm{rank}(V)\leq (\prod_id_i)^2-(\prod_id_i)+1$,  achievable for an irreducible evolution. Therefore if we allow the assumption that the system evolves unitarily, WP-G's method \emph{is} an irreducible dimension witness, but otherwise it is not.

\bibliography{dim_paper_references}
\end{document}